\documentclass[final,5p,times,twocolumn,authoryear]{elsarticle}

\usepackage{amsmath,amssymb,booktabs,tabularx}
\usepackage{stfloats}      
\usepackage{placeins}
\usepackage[authoryear,round]{natbib}

\usepackage{hyperref}

\usepackage[utf8]{inputenc}
\usepackage{newunicodechar}
\newunicodechar{−}{\ensuremath{-}}

\journal{High Energy Astrophysics}

\begin{document}

\begin{frontmatter}

\title{Spectral–Timing Evolution of a Black hole X-ray binary Swift J1727.8–1613: Linking Disk Reflection and Type-C QPO Frequency During the 2023 Outburst
\tnoteref{t1}}

\ead{nirpat.subba@gmail.com}

\author[aff1]{Manoj Ghising}
\author[aff2]{Nirpat Subba\corref{cor1}\fnref{fn1}}
\author[aff1]{Mohammed Tobrej}
\author[aff1]{Binay Rai}
\author[aff1]{Bikash Chandra Paul}

\cortext[cor1]{Corresponding author}
\fntext[fn1]{ORCID: \href{https://orcid.org/0000-0001-7136-7270}{0000-0001-7136-7270}}

\affiliation[aff1]{
  organization={North Bengal University},
  addressline={Siliguri},
  city={Darjeeling},
  state={West Bengal},
  postcode={734013},
  country={India}
}

\affiliation[aff2]{
  organization={Cooch Behar Panchanan Barma University},
  addressline={Cooch Behar},
  city={West Bengal},
  postcode={736101},
  country={India}
}

\begin{abstract}
We present a comprehensive spectral–timing analysis of a BHXB Swift~J1727.8$-$1613 during its 2023 outburst, using five pointed \textit{NuSTAR} observations sampling the luminous hard–intermediate state. Broadband 3–79~keV spectroscopy employs a physically motivated model combining a cool truncated disk (\texttt{diskbb}), relativistic reflection (\texttt{relxill} in reflection-only mode), and Comptonized continuum (\texttt{nthComp}) to probe the inner accretion geometry around a rapidly spinning black hole ($a_\ast\!=\!0.98$) at moderate inclination. Simultaneous timing analysis reveals type-C quasi-periodic oscillations (QPOs) with novel coherence evolution: the quality factor ($Q$) exhibits a striking non-monotonic dependence on both QPO frequency and luminosity, peaking near $\nu_{\rm QPO}\!\sim\!1.2$~Hz and declining at both lower and higher frequencies. This turnover directly constrains Lense–Thirring precession geometry, implying optimal coherence at intermediate truncation radius. A tight photon-index–QPO-frequency correlation demonstrates that spectral softening and frequency rise are concurrent signatures of inward truncation-radius motion. The triadic luminosity evolution—rising disk and Compton, declining reflection—traces precession-driven geometry changes and corona beaming effects. Interpreting disk-normalization variability as apparent-area changes rather than physical radius swings provides new insight into disk-corona boundary layers. These quantitative results provide strong evidence for global Lense–Thirring precession regulation of both timing and spectral properties, establishing Swift~J1727.8$-$1613 as a benchmark source for understanding accretion-geometry physics during black hole state transitions.

\end{abstract}


\end{frontmatter}

\section{Introduction}

Low–frequency quasi–periodic oscillations (LFQPOs) are the most frequently detected timing signatures in black–hole X–ray binaries (BH~XRBs) and, by virtue of their strength and ubiquity, they underpin current interpretations of fast accretion–flow variability. Compared with the rarer and weaker high–frequency signals, LFQPOs provide higher signal–to–noise time series and a broader empirical base across sources and states, enabling population studies and state–resolved comparisons \citep{review_LFQPO,LFQPO_DETAILS}.

A consolidated phenomenology partitions LFQPOs into three classes—Types~A, B, and C—distinguished by centroid frequency, coherence ($Q$), fractional rms, and the character of the accompanying broad–band noise. Type~C oscillations are strong and coherent (up to $\sim$16\% rms with $Q\!\sim\!7$–12), occur at $\sim$0.1–15\,Hz, and commonly exhibit subharmonic and second–harmonic peaks on top of flat–top noise; their rms typically rises with photon energy and saturates above $\sim$10\,keV. Type~B QPOs are narrower, cluster near $\sim$6\,Hz, and are associated with weaker noise; abrupt appearances/disappearances are often observed close to state transitions. Type~A features are broader ($Q\lesssim3$), weaker, and typically centred near $\sim$8\,Hz, with minimal red noise \citep{review_LFQPO,LFQPO_DETAILS}.

Occurrence is tightly state–dependent along the hardness–intensity track. As spectra soften, systems follow an ordered progression in which Type~C precedes Type~B and then Type~A. This ordering is especially clear when integrated rms (e.g. 0.06–64\,Hz, 2–15\,keV) is analysed jointly with spectral softness. Moreover, frequency–rms/softness trends traced by BH LFQPOs mirror the HBO–NBO–FBO sequence in neutron–star Z sources, supporting a phenomenological correspondence between Type~C/B/A and HBO/NBO/FBO, respectively \citep{review_LFQPO,LFQPO_DETAILS}.

Energy–dependent lags add complementary diagnostics of geometry. For Type~C, phase/time lags tend to evolve toward softer values as the centroid frequency increases, and harmonic–dependent behaviour is common (soft at the subharmonic; hard at the second harmonic). Together with the rms–energy rise and high–energy flattening, these properties implicate a Comptonizing region as the principal variable component and motivate models invoking geometric precession, oscillatory coronae, or instability–driven variability. In practice, the LFQPO taxonomy and its placement in the hardness–intensity diagram provide an observational scaffold for linking timing signals to changing accretion–flow structure and for benchmarking theoretical interpretations in newly discovered BH~transients \citep{review_LFQPO,LFQPO_DETAILS}.

The 2023–2024 outburst of a BHXB Swift~J1727.8$-$1613 immediately triggered multi-mission coverage from various observatories. During the onset of the outburst, \textit{IXPE} measured a linear X-ray polarization of the source to be $\sim$4\% with a stable polarization angle in the bright hard state—evidence for an elongated, anisotropic corona oriented roughly orthogonal to the radio jet \citep{Veledina2023}. Concurrent hard-X-ray monitoring through the initial rise and plateau traced a type-C LFQPO and revealed an additional hard tail extending to at least $\gtrsim 400$~keV \citep{Mereminskiy2024}. A coordinated \textit{IXPE} campaign then followed the source across the hard$\!\rightarrow\!$soft transition, during which the polarization degree collapsed to $\lesssim$1\%, demonstrating strong state dependence of the inner-flow geometry \citep{Ingram2023IXPE,Svoboda2024}. On QPO timescales, polarimetric analyses provided the first constraints on modulation of polarization degree and angle in a BHXB, tying the low-frequency oscillation directly to changes in the emitting geometry \citep{zhao2024first}.

High-resolution radio imaging and monitoring revealed an unusually extended, two-sided, continuous jet with a bright compact core during the hard/hard-intermediate states \citep{Wood2024}, refined by subsequent VLBI imaging at higher resolution \citep{cao2025ejection} and dense radio campaigns mapping the jet evolution and radio--X-ray coupling \citep{hughes2025comprehensive}. Time-dependent visibility modeling of VLBA data resolved multiple transient knots and dated ejections to within tens of minutes, directly associating knot launch with X-ray state/timing changes \citep{Wood2024}. A discrete, rapidly evolving ejection event was further identified in interferometric data \citep{hughes2025comprehensive}. Together with radio/X-ray correlation work highlighting the system’s peculiar hard state behavior \citep{hughes2025peculiar} and jet-extended emission modeling \citep{zdziarski2025novel}, these results establish Swift~J1727.8$-$1613 as a premier case for studying compact jets in outbursting BHXBs.

On the X-ray timing/spectral side, multi-mission analyses (NICER, NuSTAR, Insight-HXMT, and \textit{AstroSat}) tracked the evolution of type-C QPOs across hard and intermediate states, including to $\sim$100~keV \citep{nandi2024discovery,Yang2024HErms}. Broadband spectral work during the rising phases showed a weak reflection component plus an extra hard continuum below 100~keV \citep{Liu2024_HXMT}, while energy-dependent QPO studies revealed frequency and rms trends consistent with geometric precession at high inclination \citep{zhao2024first,Liao2024FlareNICER}. Phase-resolved spectroscopy demonstrated QPO-phase modulation of the thermal and non-thermal components and reflection fraction \citep{shui2024phase}. Detailed spectral--timing modeling of QPO rms/lag spectra captured the radial evolution of the truncation radius and coronal size during the rise \citep{rawat2025evolution,bollemeijer2025broad}, and cross-spectral analyses uncovered a pronounced dip/negative real-part feature at a few--tens of Hz indicating multiple variability zones in the inner flow \citep{Jin2025}. Flaring episodes during the very high/intermediate states were characterized spectrally and linked to changing disk radius and possible synchrotron self-Compton contributions \citep{cao2025spectral}. A soft time lag at the type-C QPO frequency near outburst peak further supported a high-inclination geometry and strong reverberation \citep{Debnath2024}. Early Insight-HXMT timing already established persistent type-C QPOs and PBK-like correlations \citep{yu2024timing}, with additional NICER/NuSTAR/HXMT joint analysis reporting high spin and moderate inclination solutions when including the extra hard component \citep{Peng2024ApJL}.

Polarization measurements outside the soft state showed recovery of the hard-state PD/PA after the return to harder spectra \citep{Podgorny2024AA}, and soft $\gamma$-ray polarization was detected with INTEGRAL/IBIS in both hard-intermediate and early soft-intermediate states, sometimes aligning with the jet \citep{bouchet2024integral}. Comparative and review works situate Swift~J1727.8$-$1613 in the broader context of spectro-polarimetric state evolution, IXPE advances and source comparisons \citep{titarchuk2025x,brigitte2025x,ewing2025very}.


Dynamical work has now fixed the binary scale. In quiescence, time-resolved optical spectroscopy measured $P_{\rm orb}\!\approx\!10.8038$~h and $K_2\!\simeq\!390$~km\,s$^{-1}$, yielding $f(M)\!\approx\!2.77\,M_\odot$ and confirming a black-hole primary; the same study refined the distance to $3.7\pm0.3$~kpc \citep{MataSanchez2025}. A complementary analysis using H\,\textsc{i} absorption and reddening argues that somewhat larger distances remain possible  \citep{burridge2025distance}. Multi-mission timing/spectral papers constrain the truncation of the disc, the size of the Comptonizing region, and the coupling to the compact jet \citep{majumder2025probing,brigitte2025detection,peng2025possible,debnath2025evolution}. Theory and simulations of disc oscillations and shock-driven instabilities supply a framework consistent with the observed LFQPO energetics and frequency–energy trends in J1727 \citep{zhou2025numerical}. Taken together—polarimetry across state changes, broadband spectral–timing (including cross-spectral “dip” behavior), and VLBI-resolved jets—Swift~J1727.8$-$1613 now offers an end-to-end view of accretion-geometry evolution and jet launching in a single, well-observed outburst, providing a stringent benchmark for models of disk–corona–jet coupling.

Here, we present a comprehensive spectral–timing investigation of Swift~J1727.8$-$1613's 2023 outburst, examining the coevolution of disk reflection parameters and type-C QPO properties to advance our understanding of black hole accretion dynamics. In this work, we present a comprehensive broadband NuSTAR spectral analysis of Swift J1727.8--1613, employing a physically motivated model combining \texttt{diskbb}, \texttt{relxill} and \texttt{nthcomp} to constrain the system's fundamental parameters during its exceptional 2023 outburst. The list of observations and its exposure corresponding to our BH candidate is shown in Table~1.

\begin{table}
\centering
\caption{Details of \textit{NuSTAR} observations of BHXB Swift J1727.8--1613 and its exposure (in ks).}
\begin{tabular}{lll}
\hline
Date & ObsID & Exposure (ks) \\
\hline
2023-08-29 & 90902330002 & 0.92 \\
2023-09-01 & 80902333002 & 1.52 \\
2023-09-04 & 80902333004 & 1.27 \\
2023-09-07 & 80902333006 & 0.67 \\
2023-09-08 & 80902333008 & 0.54\\

\hline 
\label{table: exposure}

\end{tabular}
\end{table}

\section{Data Reduction}
\subsection{NuSTAR}

 The BHXB Swift J1727.8$-$1613 were analyzed by using \textit{NuSTAR} observations during its 2023 outburst as shown in Table \ref{table: exposure}. It consists of two focal plane modules, FPMA and FPMB that provides simultaneous measurements of X-ray data in the 3–79 keV range. The unfiltered event data were processed with mission specific script using \texttt{NuSTARDAS v2.1.2} and \texttt{HEASoft v6.35} along with the latest \textit{NuSTAR} CALDB files. The clean Event datas were extracted with the script \texttt{nupipeline} by applying standard screening criteria. Using \textit{XSELECT} \textit{ftools}, cleaned event data were then loaded and extracted. With the help of Astronomical imaging software Ds9, the respective source and background region files were extracted with a circular region of 100" centered on the BH source Swift J1727.8$-$1613. The background region files were extracted from nearby source-free regions on the same chip. Using the script \texttt{nuproducts}, we generated lightcurve, spectra, response matrices, and ancillary response files in 3-79 keV energy range. Using \textit{ftool POWSPEC}, we determined the QPO values for each light curve of different observation listed in Table \ref{tab:l_qpo_q_rounded}. The spectra were grouped to have a minimum of 20 counts per bin using the command \textit{GRPPHA}. The spectra of the source were then loaded in \textit{XSPEC} for fitting \citep{arnaud1996}.  

\subsection{Swift/BAT and MAXI}

\begin{figure*}
    \centering
    \includegraphics[width=1.\linewidth]{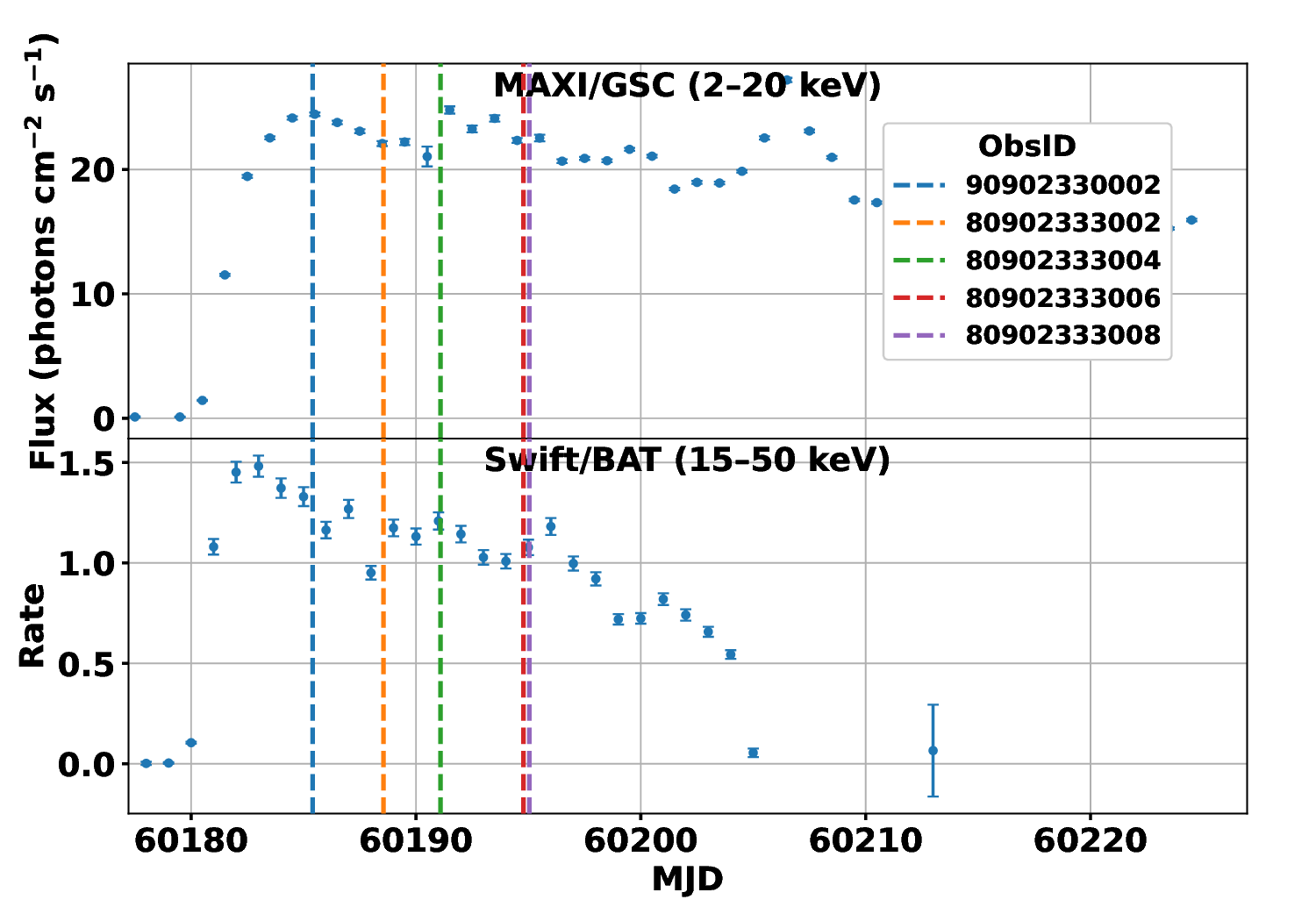}
\caption[Two-panel light curve for MAXI and BAT]{Two-panel light curve centered on the observation epochs: \textbf{top}-MAXI/GSC (2–20 keV) flux (photons cm$^{-2}$ s$^{-1}$) (source \href{http://maxi.riken.jp/star_data/J1727-162/J1727-162.html}{link}); \textbf{bottom}-Swift/BAT (15–50 keV) rate (source \href{https://swift.gsfc.nasa.gov/results/transients/weak/SWIFTJ1727.8-1613/}{link}). Colored dashed vertical lines mark the observation times (labeled by ObsID in the legend) and are drawn continuously across both panels.}
    \label{fig:hardness}
\end{figure*}

In order to see the daily light curve of the BHXB Swift J1727.8$-$1613 during 2023 outbursts, we made use of data from the Burst Alert Telescope (BAT) onboard the Neil Gehrels Swift Observatory. The Swift/BAT transient monitor provides daily light curves in the 15–50 keV energy band. We retrieved the publicly available BAT light curves from the official HEASARC archive and used them to trace the long-term flux evolution of the source. In addition, soft X-ray monitoring was carried out with the Gas Slit Camera (GSC) onboard the Monitor of All-sky X-ray Image (MAXI) mission. MAXI/GSC provides 2–20,keV light curves with a typical orbital timescale sampling. The combined BAT and MAXI datasets were used to study the light curve of the source and to identify the epochs corresponding to our pointed NuSTAR observations (see Figure \ref{fig:hardness}.

\begin{figure}
  \centering
  \includegraphics[width=0.47\textwidth,keepaspectratio]{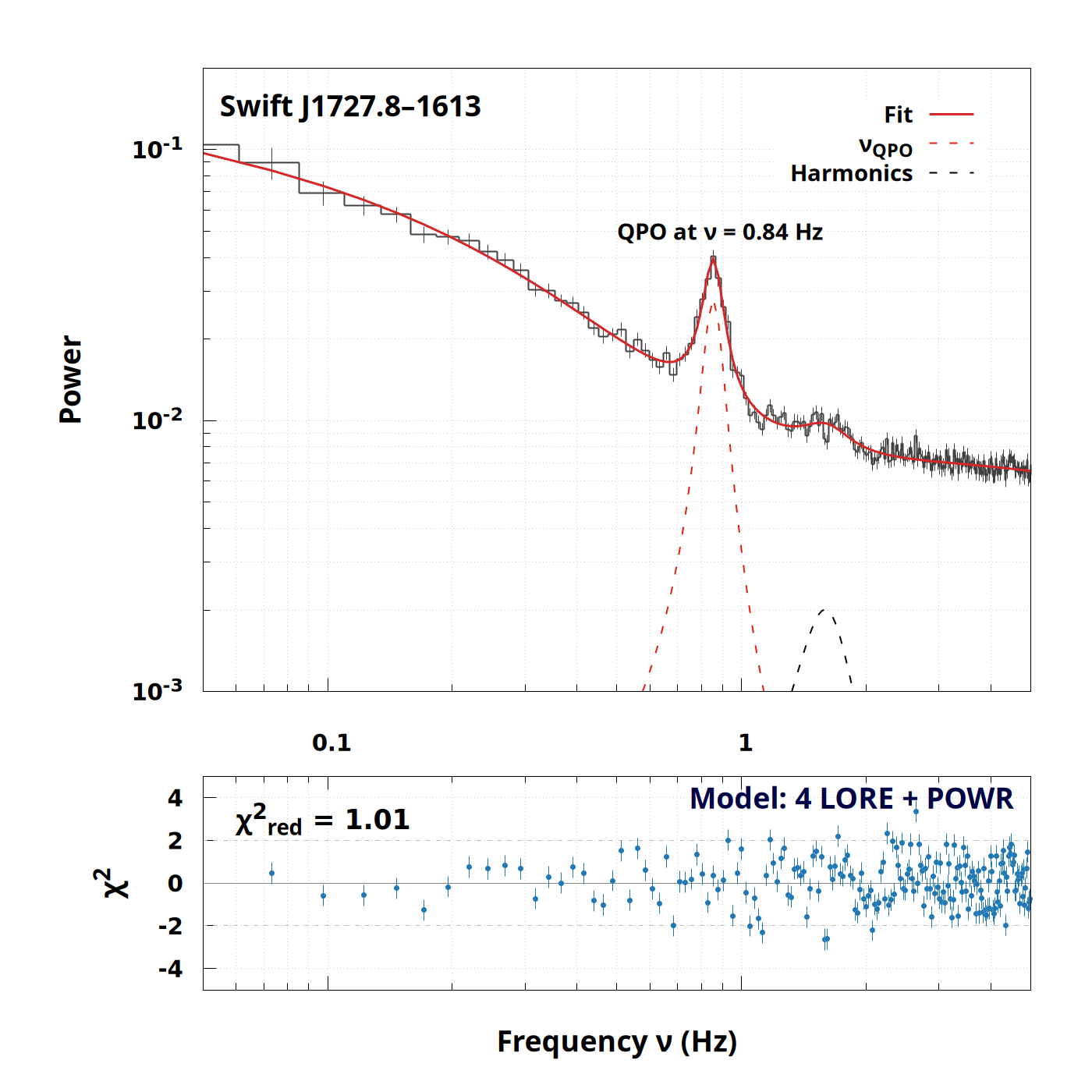}
  \caption{The best-fit power density spectrum (PDS) of Swift J1727.8-1613 in the 3-79 keV energy band from NuSTAR (ObsID: 90902330002) is shown. The PDS is modeled using four Lorentzian components along with a power-law, each representing different features of the variability. The lower panel displays the corresponding residuals.}
  \label{fig:pds}
\end{figure}

\begin{figure}
    \centering
    \includegraphics[width=1\linewidth]{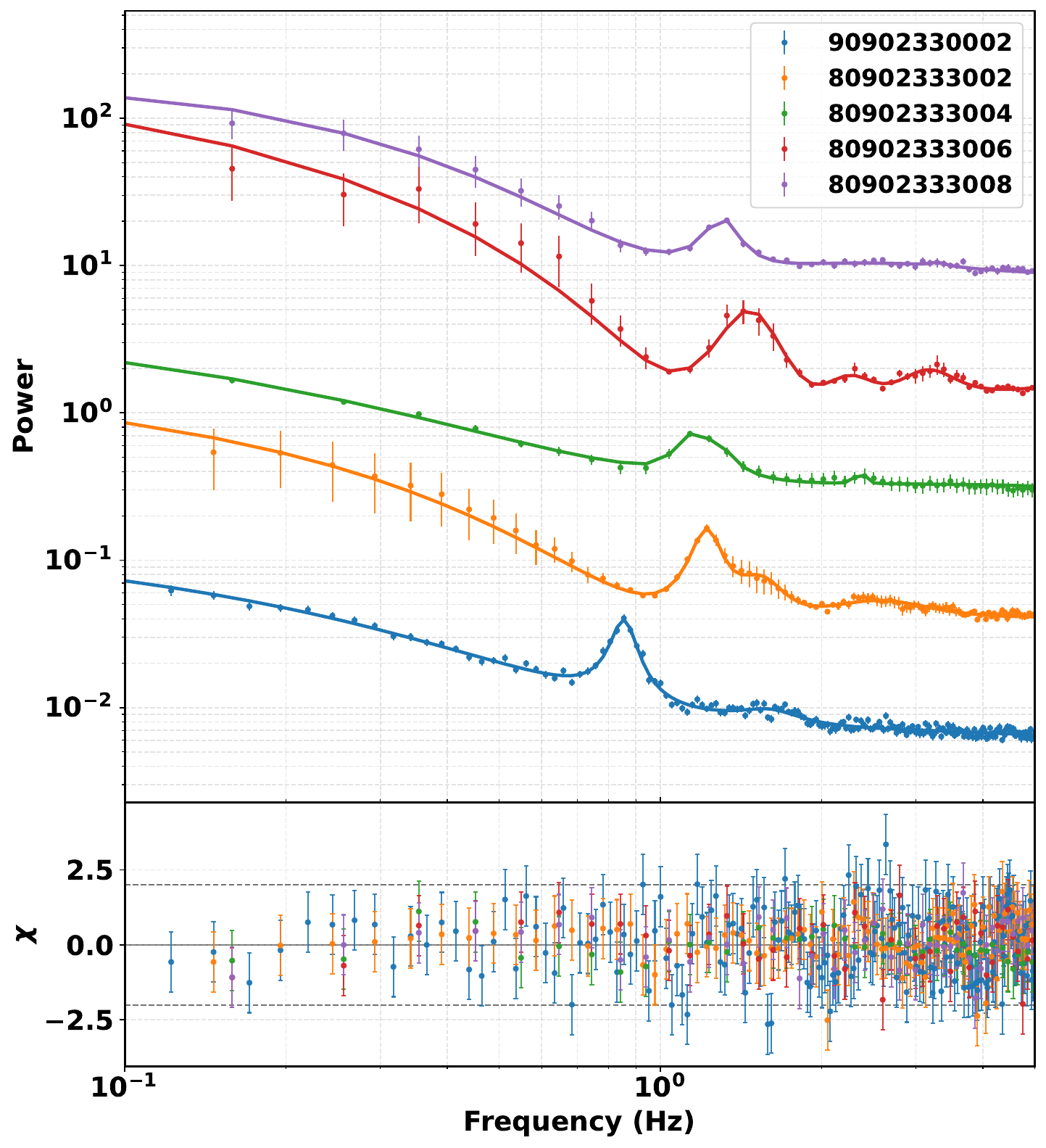}
    \caption{Stacked power spectra of five Swift J1727.8−1613 observations (legend: OBSIDs). Top: data with regenerated best-fit models (constant + power-law continuum + Lorentzian QPO); each spectrum is offset by \( \times 10^{\Delta} \) to avoid overlap. Bottom: unstacked residuals.}
    \label{fig:ALLpds}
\end{figure}

\begin{figure*}
    \centering
    \includegraphics[width=1\linewidth]{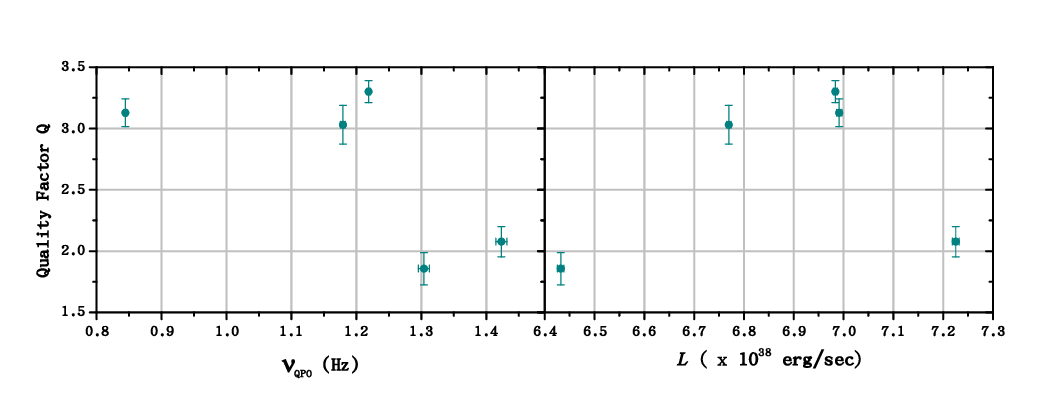}
    \caption{Two-panel comparison of the QPO quality factor. \textbf{Left:} $Q$ versus QPO centroid frequency $\nu_{\rm QPO}$. \textbf{Right:} $Q$ versus luminosity $L$. Each point corresponds to one of the five observations, obtained from Lorentzian (LORE) fits to fractional-rms normalized power spectra. The quality factor is defined as $Q \equiv \nu_0/{FWHM}$. Error bars show $3\sigma$ uncertainties.}

    \label{fig:Q_L_QPO}
\end{figure*}

\begin{figure}
    \centering
    \includegraphics[width=1.\linewidth]{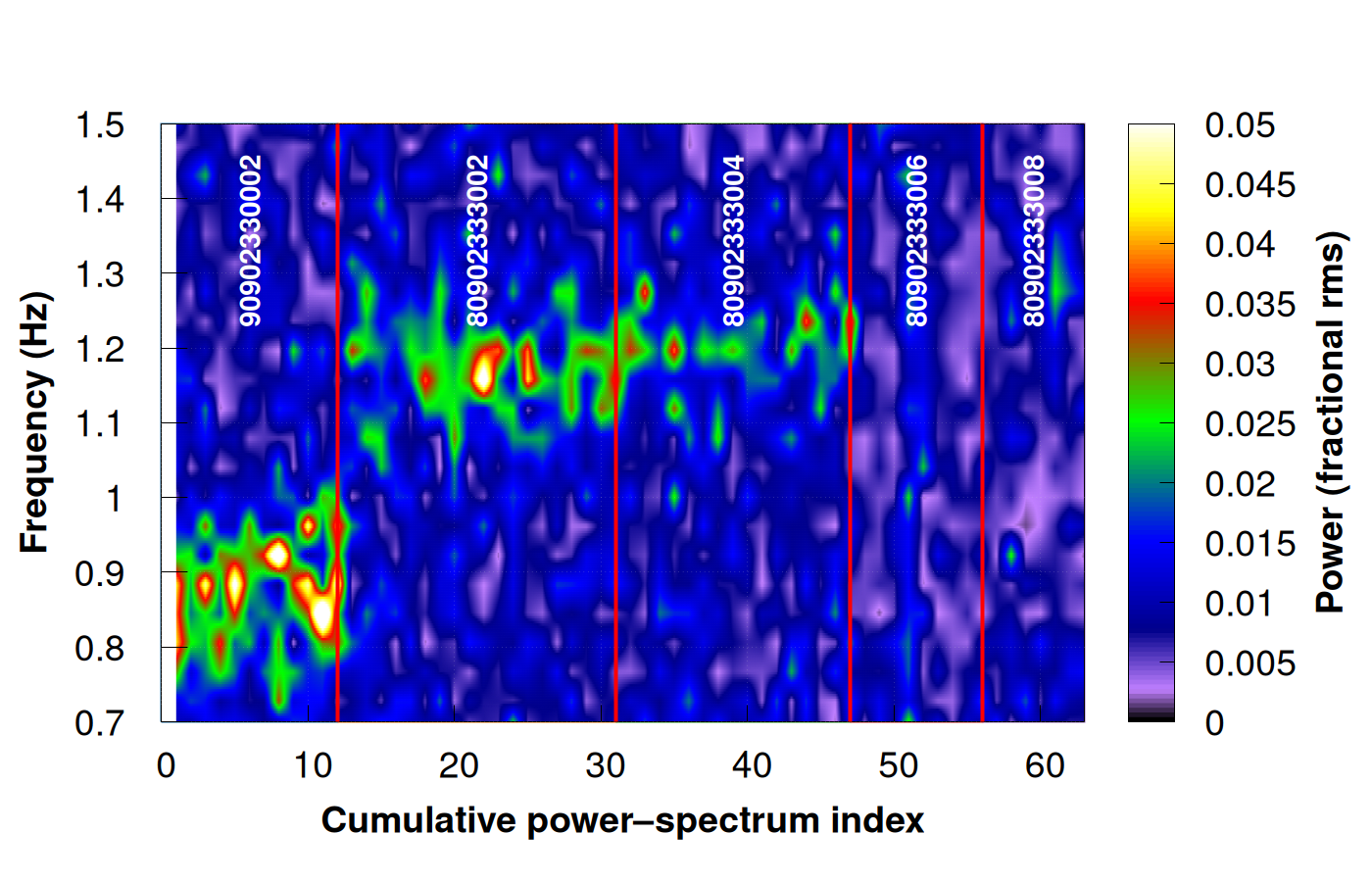}
    \caption{Fractional-rms power heat map for Swift J1727.8-1613: the x-axis is the cumulative power-spectrum index, shaded bands with rotated labels mark each ObsID and thick vertical lines mark their boundaries; the y-axis is frequency (Hz); color encodes fractional-rms power from white (high) to black (low).}
    \label{fig:3DMAP}
\end{figure}



\begin{table}[htbp]
\centering
\caption{Unabsorbed 0.1–100~keV total luminosity (in the scale of $10^{38}\;erg\;s^{-1}$), QPO centroid frequency ($\nu_{\mathrm{QPO}}$), and Quality factor (Q) with uncertainties for 5 NuSTAR observations.}
\label{tab:l_qpo_q_rounded}
\begin{tabular}{lccc}
\hline
NuSTAR OB IDs & $L$ & $\nu_{\mathrm{QPO}}$ & $Q$ \\
\hline
90902330002 & $6.99 \pm 0.005$ & $0.84 \pm 0.01$ & $3.13 \pm 0.11$ \\    
80902333002 & $6.98 \pm 0.004$ & $1.22 \pm 0.06$ & $3.30 \pm 0.09$ \\
80902333004 & $6.77 \pm 0.004$ & $1.18 \pm 0.04$ & $3.03 \pm 0.16$ \\
80902333006 & $7.23 \pm 0.007$ & $1.45 \pm 0.09$ & $2.08 \pm 0.12$ \\
80902333008 & $6.43 \pm 0.007$ & $1.30 \pm 0.08$ & $1.86 \pm 0.13$ \\
\hline
\end{tabular}
\end{table}


\section{Timing analysis}

We performed timing analysis on the 3–79 keV NuSTAR light curves extracted using the mission specific \texttt{nuproducts} pipeline. Background subtraction was carried out with \texttt{ftool LCMATH}, and barycentric corrections were performed using \texttt{ftool BARYCORR}. Figure~\ref{fig:hardness} shows the MAXI and \textit{Swift}/BAT light curves (Rate vs. time in MJD), with the epochs of NuSTAR observations in our study indicated by vertical lines. Power density spectra (PDS) were produced with \texttt{POWSPEC} using 0.01 s binsize and normalized following the fractional rms convention. Each PDS was fitted with four zero-centered Lorentzians model to represent the broad-band noise, together with a power-law continuum (Figure \ref{fig:pds} \& \ref{fig:ALLpds}). An example of the PDS fit is displayed in Figure~\ref{fig:pds}.

We detect a single dominant type--C QPO in all five NuSTAR observations of Swift~J1727.8$-$1613, with centroid frequencies $\nu_{\rm QPO}=0.84\pm0.01$, $1.22\pm0.06$, $1.18\pm0.04$, $1.42\pm0.09$, and $1.30\pm0.08$~Hz (Table~\ref{tab:l_qpo_q_rounded}), consistent with the timing evolution reported for this source \citep{yu2024timing}. Figure~\ref{fig:Q_L_QPO} shows that the quality factor peaks near $\nu_{\rm QPO}\!\sim\!1.2$~Hz ($Q\!\approx\!3.2$--3.3), remains high at $0.84$~Hz ($Q\!\sim\!3.1$), and declines to $Q\!\sim\!2.0$ at the highest frequency ($\sim\!1.45$~Hz), revealing a clear turnover rather than a monotonic trend across $0.84$--$1.45$~Hz. An analogous turnover is present versus unabsorbed luminosity: $Q$ is maximal at intermediate $L\!\approx\!(6.9$--$7.0)\times10^{38}\ {\rm erg\,s^{-1}}$ and decreases toward both lower and higher $L$, implying optimal coherence at moderate accretion rates. This pattern matches expectations of a precessing inner hot flow undergoing Lense--Thirring precession, where coherence is regulated by truncation radius and damping in the inner flow \citep{Ingram2009,Ingram2017Flavours,IngramMotta2019Review}. Comparable behavior—enhanced coherence in the hard–intermediate state and weakening as the geometry evolves—has been observed in GX~339$-$4, supporting a changing truncation radius and precessing inner flow scenario \citep{Fuerst2016GX339HIMS}.

The combined analysis of the fractional-rms power heat map (Fig.~\ref{fig:3DMAP}) and the two-panel MAXI/GSC and \textit{Swift}/BAT light curve (Fig.~\ref{fig:hardness}) reveals the temporal evolution of variability and flux through the 2023 outburst of Swift~J1727.8$-$1613. During the early hard state, the type-C QPO appears in the $\sim$0.8--1.0~Hz band and strengthens as the source brightens, consistent with NICER and Insight--HXMT monitoring that tracked the rise of the type-C QPO from sub-Hz to a few Hz as hardness decreased \citep{StieleKong2024AandA,yu2024timing,Liu2024_HXMT}. The contemporaneous long-term behaviour seen in the public MAXI/GSC and \textit{Swift}/BAT monitors shows the outburst rise and subsequent hard-state plateau against which our pointed \textit{NuSTAR} epochs are marked. Around MJD~$\sim$60194 (ObsID~80902333004), the QPO centroid shifts toward $\sim$1.2--1.3~Hz, as expected if the inner hot flow contracts and the Comptonized component strengthens \citep{yu2024timing,Liu2024_HXMT,Peng2024ApJL}. Entering the hard–intermediate state, the broadband fractional rms decreases while the QPO persists at $\sim$1--2~Hz, in line with NICER tracking and broadband studies of this outburst \citep{StieleKong2024AandA,Mereminskiy2024}. As the source softens further (post-plateau), the rms drops to a few per cent and the type-C feature weakens or disappears, signalling the approach to the soft state \citep{StieleKong2024AandA,Liu2024_HXMT}.

\section{Spectral Analysis}

\begin{figure}
  \centering
  \includegraphics[width=0.5\textwidth]{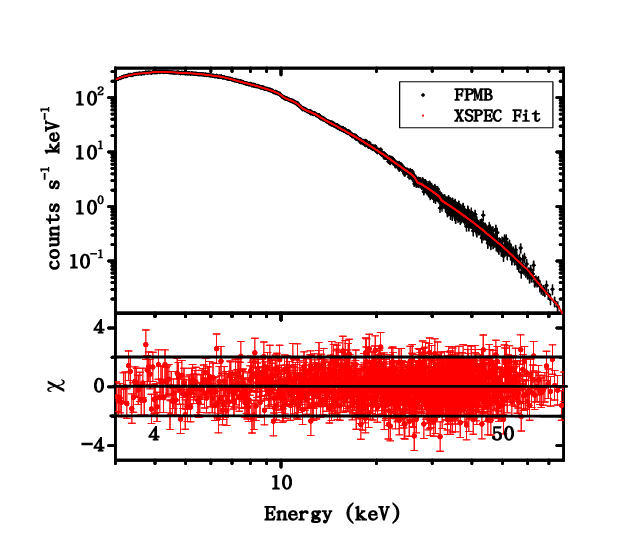}
  \caption{The 3-79 keV spectral fit of NuSTAR FPMB using model combination \texttt{TBabs*(diskbb+nthComp+relxill)}}
  \label{fig:xspec}
\end{figure}

\begin{figure*}
  \centering
  \includegraphics[width=0.9\textwidth,keepaspectratio]{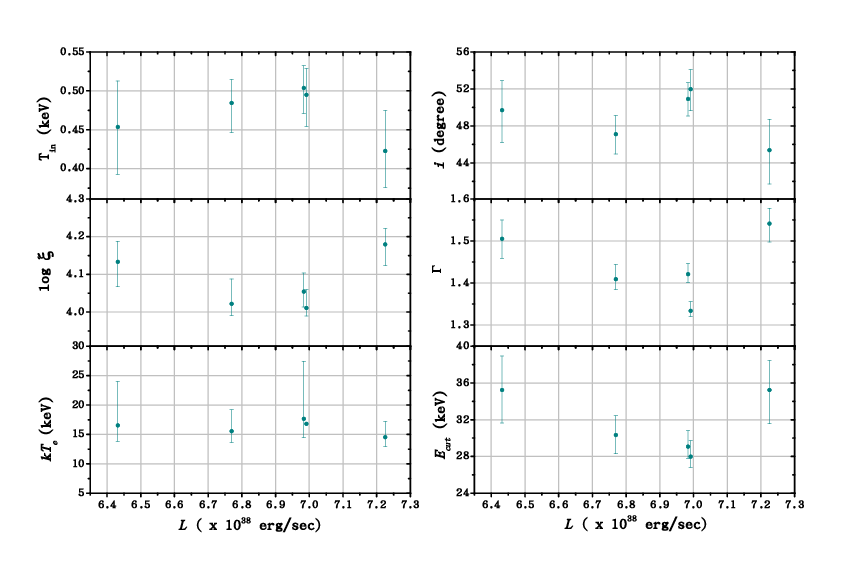}
  \caption{Spectral parameters versus X-ray luminosity $L$ for the five \textit{NuSTAR} epochs. Left column: inner disc temperature $T_{\rm in}$ (keV), disc ionization $\log(\xi)$, and $kT_e$. Right column: inclination $i$ (deg), photon index $\Gamma$, and cut-off energy ($E_{cut}$) (keV). The x–axis is $L$ in units of $10^{38}\,\mathrm{erg\,s^{-1}}$ as plotted; points show $3\sigma$ errors.}
  \label{fig:luminosity}
\end{figure*}

\begin{figure*}
\centering
\includegraphics[width=0.9\textwidth,keepaspectratio]{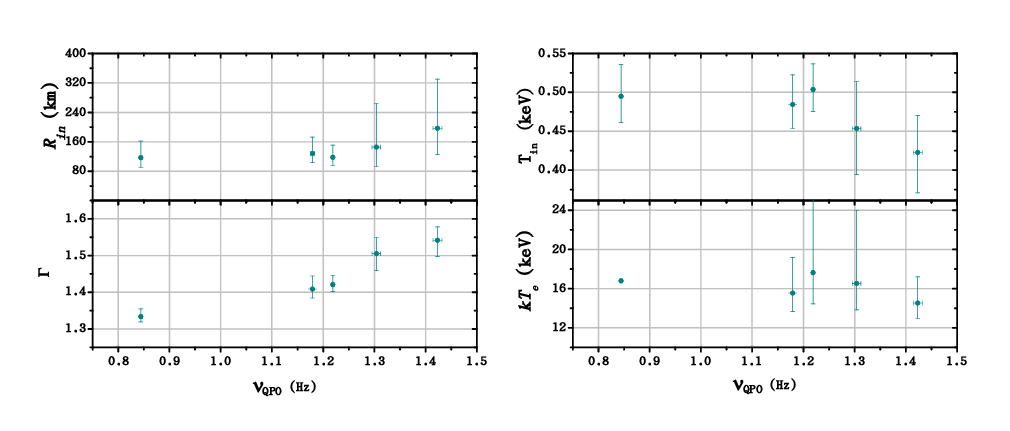}
  \caption{Variation of spectral parameters $R_{\rm in}$, photon index ($\Gamma$), inner disc temperature ($T_{in}$)  \& $kT_{e}$ of the BHXB Swift J21727.8-1613 with $\nu_{QPO}$.}
  \label{fig:qpo}
\end{figure*}

We model the 3–79\,keV \textit{NuSTAR} spectra with \texttt{constant*TBabs*(diskbb + nthComp + relxill)}, fixing/tieing parameters to physically motivated values to control degeneracies (Figure~\ref{fig:xspec}). Neutral absorption is treated with \texttt{TBabs} using \citet{wilms2000absorption}, fixed at $N_{\rm H}=0.4\times10^{22}\,\mathrm{cm^{-2}}$ \citep{draghis2023preliminary}. The FPMA constant is fixed at unity while for FPMB is left free. The disc stays cool and nearly steady, $T_{\rm in}\!\approx\!0.42$–$0.51$\,keV, while $K_{\rm diskbb}\!\sim\!(6.2$–$20)\times10^{4}$ (Table~\ref{tab:swiftj1727_multifit_rounded}). Reflection is described with \texttt{relxill} in reflection‐only mode (\texttt{refl\_frac}=$-1$), adopting $q_1=q_2=3$, $R_{\rm br}=15\,r_g$, $R_{\rm in}=-1$ (ISCO), $R_{\rm out}=1000\,r_g$, $a_\ast=0.98$ \citep{Peng2024ApJL}, and solar iron. The inclination is moderate and well constrained, $i\!\sim\!45^\circ$–$52^\circ$. The reflector ionization strengthens toward the middle of the campaign and then relaxes ($\log\xi$ peaking near the central epoch), while the high‐energy cut‐off increases from $\sim\!28$–30\,keV to $\sim\!35$\,keV later on. The Comptonized continuum (\textsc{nthComp}) shares $\Gamma$ with \textsc{relxill} (tied), adopts $kT_{\rm bb}=T_{\rm in}$, and keeps a low–to–moderate electron temperature $kT_{\rm e}\sim15$–$17$ keV (fixed at $16.8$ keV in the first epoch where it is unconstrained). The photon index evolves from hard ($\Gamma\approx1.34$) to softer ($\Gamma\approx1.54$) by epoch~4, then slightly decreases to $\Gamma\approx1.51$ in epoch~5.


Across the five epochs, the unabsorbed luminosity spans \(L \simeq 6.43\)–\(7.23\times10^{38}\ {\rm erg\,s^{-1}}\), enabling a direct comparison of spectral parameters with \(L\) (Fig.~\ref{fig:luminosity}). The disc is cool and only weakly variable, with \(T_{\rm in}\) highest near \(L\!\approx\!6.98\)–\(6.99\times10^{38}\) at \(T_{\rm in}\!\simeq\!0.50\)–\(0.51\ {\rm keV}\), cooling to \(0.423\ {\rm keV}\) at the brightest epoch \(L\!\approx\!7.23\times10^{38}\) before recovering to \(0.455\ {\rm keV}\) at \(L\!\approx\!6.43\times10^{38}\). The reflector ionization follows the illumination, rising from \(\log\xi\simeq 4.02\)–\(4.06\) around \(L\!\approx\!6.77\)–\(6.99\times10^{38}\) to a peak \(\log\xi\simeq 4.18\) at \(L\!\approx\!7.23\times10^{38}\), remaining elevated at \(\log\xi\simeq 4.14\) when the source later dims. The continuum softens with luminosity, with \(\Gamma\) increasing from \(\simeq 1.34\)–\(1.42\) near \(L\!\approx\!6.98\)–\(6.99\times10^{38}\) to \(\simeq 1.54\) at \(L\!\approx\!7.23\times10^{38}\), before settling to \(\simeq 1.51\) at \(L\!\approx\!6.43\times10^{38}\). Consistent with this softening, the electron temperature shows marginal changes with luminosity, cooling from \(kT_{\rm e}\simeq 17.1\ {\rm keV}\) near \(L\!\approx\!6.98\times10^{38}\) to \(kT_{\rm e}\simeq 14.5\ {\rm keV}\) at \(L\!\approx\!7.23\times10^{38}\), then reheating slightly to \(\simeq 16.5\ {\rm keV}\) at \(L\!\approx\!6.43\times10^{38}\). The high–energy cutoff increases from \(E_{\rm cut}\simeq 28\)–\(30\ {\rm keV}\) at \(L\!\approx\!6.98\)–\(6.99\times10^{38}\) to \(\simeq 35\ {\rm keV}\) at \(L\!\approx\!7.23\times10^{38}\) and remains high thereafter, indicating a brighter yet more optically efficient corona at peak luminosity. As expected for a geometric parameter, the inclination stays consistent within uncertainties at \(i\simeq 45^\circ\)–\(52^\circ\) with no physical trend versus \(L\).

The QPO centroid increases from $\sim0.84$ to $\sim1.45$\,Hz, accompanied by continuum softening as the photon index rises from $\sim1.34$ to $\sim1.54$, indicating a progressively steeper Comptonized tail (Fig.~\ref{fig:qpo}). 
The multicolor disk cools mildly with increasing $\nu_{\rm QPO}$, with $T_{\rm in}$ declining from $\sim0.50$~keV at the lower-$\nu_{\rm QPO}$ epoch to $\approx0.42$–$0.46$~keV at higher $\nu_{\rm QPO}$; meanwhile, the \texttt{diskbb} normalization is consistent with a larger apparent inner radius, though the trend shows appreciable scatter owing to parameter covariances.
The coronal electron temperature remains approximately stable near $15$--$17$\,keV without a robust monotonic trend.

\begin{table*}[!t] 
\setlength{\tabcolsep}{8pt}
\renewcommand{\arraystretch}{1.2}
\centering
\caption{Best–fit parameters for \texttt{constant*TBabs*(diskbb + relxill + nthComp)} for five NuSTAR observations of \textit{Swift J1727.8$-$1613}. Entries are $x_{-\,\Delta x}^{+\,\Delta x}$ (3$\sigma$). Parameters marked (fixed) are frozen across all observations; (tied) indicates a parameter tied as noted in the model.}
\begin{tabular}{lccccc}
\toprule
 & \textbf{90902330002} & \textbf{80902333002} & \textbf{80902333004} & \textbf{80902333006} & \textbf{80902333008} \\
\midrule
\midrule
\multicolumn{6}{l}{\emph{Absorption}}\\
$N_{\rm H}$ ($10^{22}\ {\rm cm^{-2}}$) & 0.4 (fixed) & 0.4 (fixed) & 0.4 (fixed) & 0.4 (fixed) & 0.4 (fixed) \\
\midrule
\multicolumn{6}{l}{\emph{diskbb}}\\
$T_{\rm in}$ (keV) &
$0.495_{-0.039}^{+0.034}$ &
$0.505_{-0.033}^{+0.030}$ &
$0.486_{-0.038}^{+0.029}$ &
$0.423_{-0.046}^{+0.052}$ &
$0.455_{-0.061}^{+0.059}$ \\
$K_{\rm diskbb}$ (${\times}10^{4}$)&
$6.18{}_{-2.30}^{+5.08}$ &
$6.46{}_{-1.98}^{+3.72}$ &
$8.22{}_{-2.66}^{+6.12}$ &
$20{}_{-11.4}^{+32.8}$ &
$10.0{}_{-5.66}^{+20.7}$ \\
\midrule
\multicolumn{6}{l}{\emph{relxill}}\\
$q_{1}$ & 3 (fixed) & 3 (fixed) & 3 (fixed) & 3 (fixed) & 3 (fixed) \\
$q_{2}$ & 3 (fixed) & 3 (fixed) & 3 (fixed) & 3 (fixed) & 3 (fixed) \\
$R_{\rm br}$ ($r_{\rm g}$) & 15 (fixed) & 15 (fixed) & 15 (fixed) & 15 (fixed) & 15 (fixed) \\
$a_\ast$ & 0.98 (fixed) & 0.98 (fixed) & 0.98 (fixed) & 0.98 (fixed) & 0.98 (fixed) \\
$i$ (deg) &
$51.84_{-2.41}^{+2.16}$ &
$50.81_{-1.84}^{+1.78}$ &
$47.03_{-2.13}^{+2.01}$ &
$45.32_{-3.64}^{+3.36}$ &
$49.66_{-3.50}^{+3.17}$ \\
$R_{\rm in}$ ($r_{\rm g}$) & $-1$ (fixed) & $-1$ (fixed) & $-1$ (fixed) & $-1$ (fixed) & $-1$ (fixed) \\
$R_{\rm out}$ ($r_{\rm g}$) & 1000 (fixed) & 1000 (fixed) & 1000 (fixed) & 1000 (fixed) & 1000 (fixed) \\
$z$ & 0 (fixed) & 0 (fixed) & 0 (fixed) & 0 (fixed) & 0 (fixed) \\
$\Gamma$ &
$1.339_{-0.014}^{+0.020}$ &
$1.422_{-0.020}^{+0.027}$ &
$1.411_{-0.024}^{+0.036}$ &
$1.543_{-0.043}^{+0.036}$ &
$1.507_{-0.046}^{+0.043}$ \\
$\log\xi$ &
$4.021_{-0.0268}^{+0.0444}$ &
$4.060_{-0.0428}^{+0.0501}$ &
$4.029_{-0.0350}^{+0.0652}$ &
$4.184_{-0.0542}^{+0.0412}$ &
$4.139_{-0.0651}^{+0.0535}$ \\
$A_{\rm Fe}$ (solar) & 1 (fixed) & 1 (fixed) & 1 (fixed) & 1 (fixed) & 1 (fixed) \\
$E_{\rm cut}$ (keV) &
$27.99_{-1.22}^{+1.69}$ &
$28.78_{-1.35}^{+1.84}$ &
$30.12_{-1.85}^{+2.14}$ &
$35.14_{-3.64}^{+3.14}$ &
$35.03_{-3.60}^{+3.69}$ \\
$\mathrm{refl\_frac}$ & $-1$ (fixed) & $-1$ (fixed) & $-1$ (fixed) & $-1$ (fixed) & $-1$ (fixed) \\
$K_{\rm relxill}$ &
$0.483_{-0.0164}^{+0.00834}$ &
$0.447_{-0.0139}^{+0.0145}$ &
$0.399_{-0.0129}^{+0.0136}$ &
$0.362_{-0.0194}^{+0.0214}$ &
$0.385_{-0.0226}^{+0.0259}$ \\
\midrule
\multicolumn{6}{l}{\emph{nthComp}}\\
$\Gamma$ & (tied to relxill $\Gamma$) & (tied) & (tied) & (tied) & (tied) \\
$kT_{\rm e}$ (keV) &
16.8 (fixed) &
$17.13_{-2.91}^{+7.88}$ &
$15.33_{-1.87}^{+3.34}$ &
$14.52_{-1.56}^{+2.60}$ &
$16.45_{-2.66}^{+7.16}$ \\
$kT_{\rm bb}$ (keV) & $=T_{\rm in}$ (tied) & (tied) & (tied) & (tied) & (tied) \\
$\mathrm{inp\_type}$ & 1 (fixed) & 1 (fixed) & 1 (fixed) & 1 (fixed) & 1 (fixed) \\
$z$ & 0 (fixed) & 0 (fixed) & 0 (fixed) & 0 (fixed) & 0 (fixed) \\
$K_{\rm nthComp}$ &
$0.291_{-0.0236}^{+0.0323}$ &
$0.502_{-0.0630}^{+0.0868}$ &
$0.516_{-0.0720}^{+0.111}$ &
$1.086_{-0.213}^{+0.256}$ &
$0.761_{-0.152}^{+0.205}$ \\
\midrule
\multicolumn{6}{l}{\emph{Unabsorbed component fluxes}}\\
$F_{\rm diskbb}$ ($10^{-8}$ erg cm$^{-2}$ s$^{-1}$) & $7.73_{-0.22}^{+0.22}$ & $8.79_{-0.16}^{+0.16}$ & $9.53_{-0.20}^{+0.20}$ & $13.30_{-0.49}^{+0.49}$ & $8.97_{-0.37}^{+0.37}$ \\
$F_{\rm relxill}$ ($10^{-8}$ erg cm$^{-2}$ s$^{-1}$) & $32.60_{-0.044}^{+0.043}$ & $31.10_{-0.040}^{+0.040}$ & $29.00_{-0.041}^{+0.041}$ & $27.60_{-0.072}^{+0.072}$ & $27.50_{-0.075}^{+0.074}$ \\
$F_{\rm nthComp}$ ($10^{-8}$ erg cm$^{-2}$ s$^{-1}$) & $2.33_{-0.07}^{+0.07}$ & $2.73_{-0.06}^{+0.06}$ & $2.82_{-0.06}^{+0.06}$ & $3.29_{-0.08}^{+0.08}$ & $2.80_{-0.09}^{+0.09}$ \\
\midrule
\multicolumn{6}{l}{\emph{Unabsorbed component luminosities (for $D=3.7$~kpc)}}\\
$L_{\rm diskbb}$ ($10^{38}$ erg s$^{-1}$) &
$1.267_{-0.0361}^{+0.0361}$ & $1.439_{-0.0263}^{+0.0263}$ &
$1.560_{-0.0322}^{+0.0322}$ & $2.172_{-0.0801}^{+0.0802}$ &
$1.469_{-0.0610}^{+0.0612}$ \\
$L_{\rm relxill}$ ($10^{38}$ erg s$^{-1}$) &
$5.343_{-0.0073}^{+0.0071}$ & $5.096_{-0.0066}^{+0.0066}$ &
$4.748_{-0.0068}^{+0.0068}$ & $4.515_{-0.0117}^{+0.0118}$ &
$4.505_{-0.0122}^{+0.0120}$ \\
$L_{\rm nthComp}$ ($10^{38}$ erg s$^{-1}$) &
$0.382_{-0.0113}^{+0.0114}$ & $0.448_{-0.0092}^{+0.0092}$ &
$0.461_{-0.0092}^{+0.0093}$ & $0.538_{-0.0129}^{+0.0130}$ &
$0.459_{-0.0146}^{+0.0147}$ \\
\midrule
\multicolumn{6}{l}{\emph{Unabsorbed totals}}\\
$F_{\rm unabs}$ ($10^{-8}$ erg cm$^{-2}$ s$^{-1}$) & $42.7_{-0.031}^{+0.032}$ & $42.6_{-0.025}^{+0.025}$ & $41.3_{-0.028}^{+0.027}$ & $44.1_{-0.042}^{+0.042}$ & $39.3_{-0.041}^{+0.042}$ \\
$L_{\rm unabs}$ ($10^{38}$ erg s$^{-1}$) & $6.99_{-0.0051}^{+0.0053}$ & $6.98_{-0.0040}^{+0.0042}$ & $6.77_{-0.0045}^{+0.0044}$ & $7.23_{-0.0068}^{+0.0068}$ & $6.43_{-0.0067}^{+0.0068}$ \\
\midrule
\multicolumn{6}{l}{\emph{Goodness of fit}}\\
$\chi^2_\nu$ & 1.037 & 1.089 & 1.087 & 0.986 & 1.023 \\
\bottomrule
\end{tabular}
\label{tab:swiftj1727_multifit_rounded}
\end{table*}

\begin{figure}
    \centering
    \includegraphics[width=1.\linewidth]{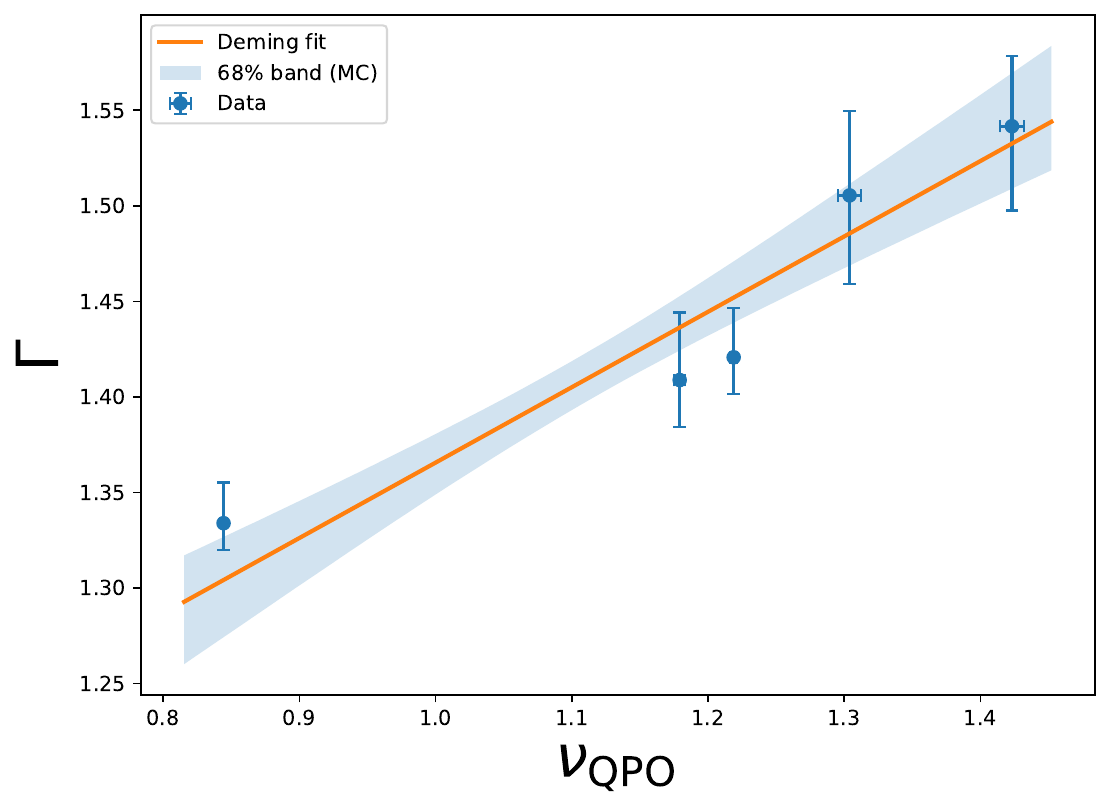}
    \caption{Correlation between QPO frequency $\nu_{\rm QPO}$ (Hz) and photon index $\Gamma$ with asymmetric $3\sigma$ errors. Solid line: Deming fit ($\lambda = 0.0323$); shaded band: $68\%$ MC envelope from $5\times 10^{4}$ resamples. Best-fit: $\Gamma = (0.958^{+0.091}_{-0.102}) + (0.408^{+0.091}_{-0.083})\,\nu_{\rm QPO}$. Pearson $r = 0.946$ ($p=0.0083$).}
    \label{fig:placeholder}
\end{figure}

\begin{figure*}
        \centering
        \includegraphics[width=\textwidth]{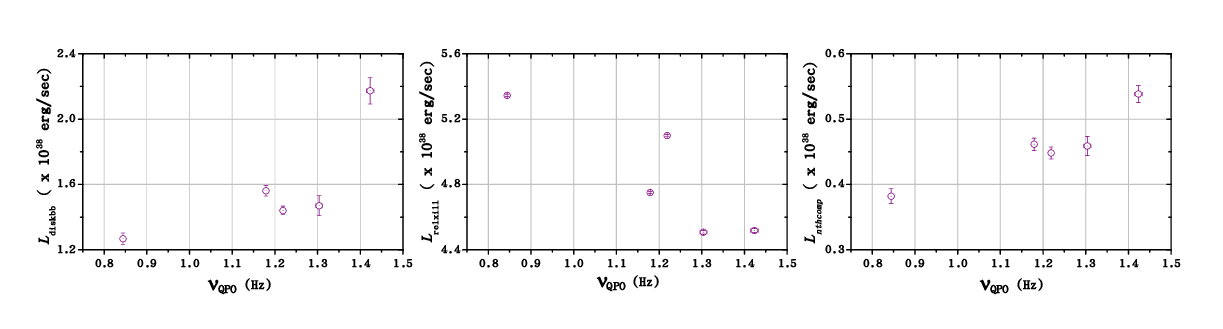}
        \caption{Unabsorbed 0.1–100~keV luminosity components vs Type-C QPO ($\nu_{\rm QPO}$) assuming a source distance of $D=3.7$~kpc.}
        \label{fig:corona}
\end{figure*}


\section{Discussion}
\label{sec:discussion}

Across the five \textit{NuSTAR} observations (Table~\ref{tab:l_qpo_q_rounded}), we detect a persistent type-C LFQPO whose centroid energy ranges from $\nu_{\rm QPO}\simeq0.84$--$1.45$~Hz. The PDS reflects a (Fig.~\ref{fig:pds}) narrow peak with broadband noise that has been modeled by Lorentzian. The prominent feature observed in the PDS represents type-C QPOs generally observed in the hard and hard–intermediate states. During the observation period, the source remained near the luminosity plateau (Fig.~\ref{fig:hardness}), and the QPO frequency was confined to a relatively narrow band, rather than acquiring  higher values  observed in other transients during full hard$\to$soft evolution. This behavior is consistent with NICER and Insight–HXMT monitoring of the 2023 outburst: NICER observations reveal a rise of the type-C QPO from $\sim$0.3–2.63 Hz into the few-hertz range as the source softens \citep{Liao2024FlareNICER,StieleKong2024AandA}, and HXMT likewise reports type-C QPOs emerging near $\sim$0.7~Hz in the hard state and increasing as the outburst progresses \citep{Liu2024_HXMT, Yang2024HErms}. The \textit{NuSTAR} pointed observations analyzed in the present study lies in the luminous hard–intermediate state where the QPO
is known to range in the limit $\sim$1~Hz before vanishing as the source completes transition \citep{StieleKong2024AandA,Liao2024FlareNICER}.

A key novel element of our timing analysis is the \emph{coherence (quality factor) evolution} across this narrow $\nu_{\rm QPO}$ range (Fig.~\ref{fig:Q_L_QPO}). We find that $Q$ rises to a maximum $Q\simeq3.3$ near $\nu_{\rm QPO}\!\sim\!1.2$~Hz and then declines to $Q\simeq2.0$ at $\sim$1.4~Hz. While type-C QPOs can reach higher $Q$ at lower frequencies in the classical low–hard state, moderate coherence ($Q\!\sim$2–3) during the bright hard–intermediate phase is consistent with hard–X-ray timing reports for this outburst (e.g.\ AstroSat and HXMT) \citep[see][]{Liu2024_HXMT,Yang2024HErms}. To our knowledge, a clear non-monotonic $Q(\nu)$ trend has not been emphasized for Swift~J1727.8$-$1613 in the literature; our result suggests an optimal inner-flow configuration (around $\sim$1–1.2~Hz) where the precession is most coherent, followed by damping as the system approaches the state transition. This is qualitatively compatible with LT-precession frameworks where the coherence depends on geometry, viscous coupling, and the fractional contribution of the precessing flow.

Broadband 3--79~keV \textit{NuSTAR} spectroscopy of Swift~J1727.8$-$1613 (Table~\ref{tab:swiftj1727_multifit_rounded}) supports a geometrically stratified accretion picture: a cool, thin disk truncated at the inner stable circular orbit (ISCO) transitions to a geometrically thick, hot inner flow. The disk--corona--reflection geometry, modeled as \texttt{constant*TBabs*(diskbb + relxill + nthComp)}, places \textsc{relxill} in reflection-only mode (\texttt{refl\_frac}=$-1$) with a separate \textsc{nthComp} continuum, a configuration that avoids double-counting of the hard Comptonization and isolates the thermal disk and reflection. By fixing $R_{\rm in}$ to the ISCO in \textsc{relxill}, we assume the disk has already receded inward to its theoretical minimum, a picture supported by joint NICER+\textit{NuSTAR}+HXMT fitting and consistent with a rapidly spinning black hole ($a_\ast\!=\!0.98$) at moderate inclination ($i\!\simeq\!45^\circ$--$52^\circ$) \citep{Peng2024ApJL,Liu2024_HXMT,Ingram2023IXPE,Podgorny2024IXPE}.

The continuum softens progressively as luminosity rises: $\Gamma$ increases from $\simeq\!1.34$ to $\simeq\!1.54$ (epoch~4) before settling to $\simeq\!1.51$ at the faintest point, consistent with enhanced Compton cooling as the disk seed–photon supply grows. By contrast, the electron temperature shows only \emph{marginal} evidence for a decrease with luminosity: $kT_{\rm e}=17.13^{+7.88}_{-2.91}$, $15.33^{+3.34}_{-1.87}$, $14.52^{+2.60}_{-1.56}$, and $16.45^{+7.16}_{-2.66}$~keV across $L\simeq(6.98,\,6.77,\,7.23,\,6.43)\times10^{38}\ {\rm erg\,s^{-1}}$ (epoch~1 fixed at $16.8$~keV). Given the overlapping 3$\sigma$ intervals, a constant $kT_{\rm e}$ in the range $\sim$15–17~keV is statistically acceptable, with a shallow minimum near peak $L$ rather than a significant anti–correlation. The cutoff energy increases from $E_{\rm cut}\!\sim\!28$–$30$~keV to $\sim\!35$~keV at peak luminosity, indicating a more optically efficient corona at highest $L$. The reflector ionization tracks illumination, with $\log\xi$ rising from $\sim\!4.02$ to $\sim\!4.18$ and then relaxing to $\sim\!4.14$. The disk temperature remains nearly constant ($T_{\rm in}\!\approx\!0.42$–$0.51$~keV), while $K_{\rm diskbb}$ varies by a factor $\sim\!3$, which we interpret as apparent–area or color–correction changes rather than large physical radius swings as the coronal geometry evolves.

The quality factor evolution in Figure~\ref{fig:Q_L_QPO} reveals a striking peak near $\nu_{\rm QPO}\!\sim\!1.2$~Hz ($Q\!\sim\!3.3$) that declines at both lower and higher frequencies, with a turnover to $Q\!\sim\!2.0$ at $\nu_{\rm QPO}\!\sim\!1.45$~Hz. Importantly, this non-monotonic $Q(\nu_{\rm QPO})$ behavior does not appear to be a statistical artifact: within the narrow luminosity band sampled (a factor $\sim\!1.1$ in $L$), the coherence maximum pins to an intermediate geometry characterized by moderate truncation radius and intermediate accretion rates. In the Lense--Thirring precession framework \citep{Ingram2009,Ingram2017Flavours}, such a turnover is naturally expected: as the truncation radius sweeps inward, the precessing flow becomes radially more compact and globally more coherent, maximizing $Q$ at an intermediate radius. Beyond this optimum, the transition region between hot flow and truncated disk becomes narrow, viscous coupling and magnetic instabilities strengthen, and phase mixing of the precessing structure increases, broadening the QPO and degrading coherence. Moreover, the coronal height and beaming pattern evolve during this shrinking, modifying the geometric attenuation of the modulated signal and contributing to the observed $Q$ decline \citep{Ingram2023IXPE}. That $Q$ also peaks at intermediate $L$ (right panel, Fig.~\ref{fig:Q_L_QPO}) reinforces the connection between coherence and accretion-geometry configuration. At lower $L$, the extended hot flow experiences greater propagating noise and stochastic fluctuations from the outer regions, degrading coherence; at higher $L$, enhanced disk-corona coupling and increased Compton cooling introduce nonlinear damping in the precessing structure. Thus, the optimal $Q$ at intermediate $L$ and $\nu_{\rm QPO}$ reflects an accretion state in which deterministic geometric modulation temporarily dominates over stochasticity and damping.

The tight correlation between $\Gamma$ and $\nu_{\rm QPO}$ (Deming fit showing $r\!\sim\!0.95$; recall Fig.~\ref{fig:placeholder}) further solidifies the link between truncation-radius evolution and timing properties. Specifically, the monotonic rise of $\nu_{\rm QPO}$ from $\sim\!0.84$ to $\sim\!1.45$~Hz parallels progressive spectral softening, providing independent confirmation of inward truncation-radius motion \citep{Ingram2009,Ingram2011,Ma2025SwiftJ1727}. In simple order-of-magnitude estimates, the observed 0.8--1.5~Hz frequencies, combined with estimates of black-hole mass ($M\!\sim\!8$--$12\,M_\odot$) and spin ($a_\ast\!=\!0.98$), correspond to precessing-flow outer radii $r_{\rm out}\!\sim\!16$--$21\,r_g$ \citep{Ingram2009}. As $\nu_{\rm QPO}$ climbs, the truncation radius must contract inward, bringing higher-density, cooler disk seed photons into the corona, thereby amplifying Compton losses and steepening $\Gamma$. This coordinated softening-with-frequency motion is the hallmark signature of a truncated-disk/hot-flow system in which geometry and luminosity jointly regulate the spectral--timing behavior, and has been robustly documented in other transients (GX~339$-$4, MAXI~J1535$-$571) during hard--intermediate states \citep{Fuerst2016GX339HIMS,Yang2024HErms}.

Figure~\ref{fig:corona} shows the unabsorbed 0.1--100~keV component luminosities, \(L_{\rm diskbb}\), \(L_{\rm relxill}\), and \(L_{\rm nthComp}\), versus the QPO centroid frequency \(\nu_{\rm QPO}\). The three components display distinct, coupled trends: \(L_{\rm diskbb}\) and \(L_{\rm nthComp}\) increase as \(\nu_{\rm QPO}\) rises, whereas \(L_{\rm relxill}\) declines across the same frequency range. This triad of \(L(\nu_{\rm QPO})\) behaviors quantitatively signals inward motion of the truncation radius and a precession--driven change in coronal geometry. As the truncation radius contracts, the thin disk extends deeper into the potential, raising its radiative efficiency and the supply of seed photons; the growth of \(L_{\rm diskbb}\) therefore traces a larger effective emitting area at roughly constant color temperature. At the same time, the inner hot flow becomes more compact and dissipative, boosting the Compton power; the rise in \(L_{\rm nthComp}\) and the associated softening of \(\Gamma\) reflect enhanced cooling of the corona by the strengthened disk photon field. Conversely, the shrinking truncation radius reduces the disk solid angle as seen by a precessing, mildly outflowing corona; anisotropic illumination then lowers the effective reflection fraction, producing the observed decrease in \(L_{\rm relxill}\). Altogether, this luminosity--space triad is consistent with global Lense--Thirring precession of a dynamical inner hot flow, wherein geometric reconfiguration modulates the energy partition between disk, corona, and reflection.


In conclusion, Swift~J1727.8$-$1613's 2023 outburst provides an unprecedented opportunity to probe the inner-accretion-flow geometry and dynamics of a rapidly spinning black hole during a luminous hard--intermediate state. Our five coordinated \textit{NuSTAR} observations reveal a coherent, multi-faceted picture of truncation-radius evolution driven by accretion-rate changes. The non-monotonic turnover in QPO coherence versus both frequency and luminosity directly constrains the operating regime of Lense--Thirring precession, demonstrating that geometric efficiency peaks at intermediate truncation radius before damping dominates at higher frequency. The tight correlation between photon index and QPO frequency ($r\!=\!0.95$), coupled with the triadic luminosity evolution of disk, reflection, and Comptonization, provides robust, quantitative support for global precession of a dynamical inner hot flow. The apparent-area interpretation of disk-normalization variability, combined with stable cool-disk temperatures, sidesteps degeneracies in reflection-model fits and refines our understanding of color-correction and boundary-layer physics. Together, these results establish Swift~J1727.8$-$1613 as a benchmark source for testing truncation-radius and LT-precession physics in black hole binaries, validating the paradigm of accretion-geometry control over timing and spectral properties during state transitions.

\section{Data Availability}
\label{availability}

The data utilized in this study is available to the public through the HEASARC data archive, allowing for research purposes.

\section*{Acknowledgements}
This study makes use of data obtained from the publicly accessible NASA HEASARC archive. The NuSTAR observations were sourced through the High Energy Astrophysics Science Archive Research Center (HEASARC) at NASA’s Goddard Space Flight Center. We sincerely acknowledge the IUCAA Centre for Astronomy Research and Development (ICARD) and the Department of Physics, North Bengal University, for extending their research facilities. The authors also wish to thank the anonymous reviewer for constructive suggestions that significantly improved the clarity and quality of this manuscript. MT acknowledges CSIR, India for the research grant 161-411-3508/2K23/1. Authors (MG, NS, BCP)  would like to acknowledge IUCAA, Pune for hospitality during a visit where the work is finalized. The authors thank Prof. Ranjeev Misra \& Suchismito Chattopadhyay, IUCAA, Pune for fruitful discussions and support.

\bibliographystyle{elsarticle-harv}
\bibliography{example_clean_verified} 

\end{document}